\def\im{\,{\rm Im}\,}
\def\re{\,{\rm Re}\,}
\def\ibar{\bar\imath}
\newfont{\gothic}{eufm10}
\newfont{\blackboard}{msbm10}
\def\bbf#1{\mbox{\blackboard #1}}
\def\goth#1{\mbox{\gothic #1}}
\def\Fd{{\bbf D}}
\documentstyle{article}
\begin{document}
\def\skipu{12pt}
\parskip=\skipu
\newtheorem{theorem}{Theorem}
\newtheorem{definition}{Definition}
\newtheorem{corollary}{Corollary}
\title{Non-linear Schr\"odinger Equations, \\ Separation and Symmetry }
\author{George Svetlichny \\
Departamento de Matem\'atica\\
Pontif\'{\i}cia Universidade Cat\'olica\\
Rio de Janeiro\\
Brazil}
\maketitle

\begin{abstract}
We investigate the symmetry properties of hierarchies of non-linear
Schr\"odinger equations, introduced in \cite{GS:separation}, which
describe non-interacting systems in which tensor product wave-functions
evolve by independent evolution of the factors (the separation
property). We show that there are obstructions to lifting symmetries
existing at a certain number of particles to higher numbers. Such
obstructions vanish for particles without internal degrees of freedom
and the usual space-time symmetries. For particles with internal degrees
of freedom, such as spin, these obstructions are present and their
circumvention requires a choice of a new term in the equation for each
particle number. A Lie-algebra approach for non-linear theories is
developed.
\end{abstract}

\section{Introduction}
Reference \cite{GS:separation} investigates hierarchies of non-linear
Schr\"odinger equations focusing on the property that is called {\em
separation\/} which means that tensor product functions evolve by
separate evolution of the factors.  Such systems are considered to be
non-interacting. In this paper we continue the investigation of
evolutions satisfying the separation property focusing now on the
questions of symmetries and on further mathematical
properties of these hierarchies in general, providing thus a series of
basic results necessary for the exploration of theories of this type.
In particular we develop a Lie algebra approach to infinitesimal
symmetries, adequately modified to account for non-linearities. Though
this
paper is a continuation of \cite{GS:separation},  we've made it
self contained.

It was
pointed out in \cite{GS:separation} that the theories here considered
exhibit two new physical aspects not present in linear theories. One is
the possibility of multi-particle terms in the evolution that vanish
whenever the wave-function is a tensor product. These terms introduce
truly new multi-particle effect that can only be seen in correlated
systems and that are not apparent in systems of fewer number of
particles. The other is the existence of two new universal physical
constants with the dimension of energy which describe the effect that
changing the overall phase of the initial data has on the evolution of a
wave-function. The present paper adds to such differences by showing
that  symmetries existing in non-linear equations up to a certain number
of particles do not in general persist at a greater number of particles.
In other words, symmetries can be broken by mere particle number
increase. Theorem~\ref{thm:obsatl} in Section~\ref{sec:Laf} provides the
general result allowing us to calculate some of the obstructions to
extending infinitesimal symmetries to a greater number of particles. In
the same section, Corollary~\ref{cor:onepsymlift} shows that
one-particle infinitesimal symmetries lift to any number of particles if
and only if they lift to two particles, and
Corollary~\ref{cor:addgensymm} shows that if at particle-number \(\ell\)
one introduces a new multi-particle effect of the type envisaged by
these theories, and if at \(\ell\) particles this effect does not break
an infinitesimal symmetry lifted from a one-particle symmetry, then the
symmetry is not broken at any higher number of particles if and only if
it is not broken for \(\ell +1\) particles. Theorem~\ref{thm:freelift}
in Section~\ref{sec:spacetime} then shows that, remarkably enough, for
particles without internal degrees of freedom, these obstructions at the
next particle number vanish for the usual infinitesimal space-time
symmetries. Theories of particles with internal degrees of freedom do
not escape such obstructions, which poses a problem. In particular, for
the case of spin greater than zero, one must either face the possibility
that rotation invariance be broken at a certain particle number, or else
one must introduce some new physical principle to systematically provide
the proper rotationally covariant multi-particle equation for each
number of particles.

Non-linear theories, as mentioned in \cite{GS:separation}, may in the
end be found untenable. Their detailed study exposes the difficulties
they must face, such as the symmetry lifting obstructions described in
this paper. One can through this, even if one ultimately discards such
theories, achieve a deeper understanding of why the quantum world as
seen in the laboratory is linear to such a high degree.

\section{General conventions}
When we deal with complex functions defined on subsets of the complex
plane we shall not assume any analyticity properties unless explicitly
stated. Although it is traditional in such cases to write
\(f(z,\bar z)\) instead of \(f(z)\) we shall write just \(f(z)\). Our
linear spaces will be considered real even though they may be ostensibly
complex, such as spaces of complex-values functions. In such cases, it is
the real structure of the conventional complex space which is used. By
an ``operator'' we shall mean a map \(F\) from some domain in a real
linear space \({\bf V}\) with values in some other real linear space
\({\bf }\) and we write \(F:{\bf V} \to {\bf W}\) omitting any explicit
mention
of the domain. No linearity is implied by the term ``operator''. An
operator \(F\) applied to a vector \(\phi\) shall be denoted either by
\(F\phi\) or \(F(\phi)\), the choice  being dictated by clarity and
simplicity of expression. For operators defined between spaces of
complex functions we shall use the term ``linear'' to mean {\em
complex-linear} and so the term ``real-linear'' will never be
abbreviated when meant.  For an operator \(F\) of the type we consider,
acting on a complex-valued function \(\phi\), one traditionally  would
write \(F(\phi,\bar\phi)\)  to express that complex conjugation is
allowed, but we again adopt the simpler form \(F(\phi)\). Given an
operator \(F:{\bf V}\to {\bf W}\) we shall denote by \(\Fd F(\phi)\) its
Fr\'echet derivative at \(\phi\). This is a real-linear map from \({\bf
V}\) to \({\bf W}\) that satisfies \(F(\phi +  \eta) = F(\phi) + \Fd
F(\phi)\cdot \eta + o(\eta)\). This of course means that the spaces must
have appropriate topologies for this to be well defined. If \(G: {\bf V}
\to {\bf V}\) is another map  we shall denote by \(\Fd F\cdot G\) the
operator that maps \(\phi\) to \(\Fd F(\phi)\cdot G(\phi)\).    The
product \(F,G \mapsto \Fd F \cdot G\) is real-bilinear but not
associative. However, for \({\bf W}= {\bf V}\) the commutator
\([F,G] = \Fd F \cdot G - \Fd G \cdot F\) is a Lie
bracket and is seen to be
the usual Lie bracket of \(F\) with \(G\) considering these
as vector fields on \({\bf V}\). Given \(r\) functions \(\alpha_j : {\bf X}_j
\to {\bbf C}\), \(j=1,\dots,r\) defined on some sets \({\bf X}_j\), we denote
by \(\alpha_1 \cdot \alpha_2 \cdots \alpha_r\) their tensor product
defined in the usual way on the cartesian product of the domains \({\bf X}_1
\times {\bf X}_2\times \cdots\times {\bf X}_r\). A particular case of this is when
\({\bf X}_j = {\bf X}^{n_j}\) in which case we can interpret
\(\alpha_1\cdots\alpha_r\) as being defined on \({\bf X}^n\) where \(n =
n_1+\cdots+n_r\).

\section{Hierarchies of equations and operators}
Reference~\cite{GS:separation} provides the situation that motivated the
present investigation. There one has
a hierarchy of multi-particle evolution equations, one for
each number of particles of designated species. Species merely
differentiate one-particle evolutions, the particles are otherwise
distinguishable. In such a context, for an \(n\)-tuple of species
\(s=(s_1,\dots,s_n)\) and for an \(n\)-tuple of particle positions \(x =
(x_1,\dots,x_n)\) where \(x_j = (x_j^1,\dots,x_j^d)\) are the position
coordinates of the \(j\)-th particle in \(d\)-dimensional space, we have
an \(n\)-particle probability amplitude \(\psi^{(s)}(t, x) =
\psi^{(s)}(t, x_1,\dots,x_n)\) whose square modulus \(|\psi^{(s)}(t,
x)|^2\) is the joint probability density of finding at time \(t\) the
particles at the corresponding positions \(x_j\). These amplitudes obey
a set of evolution equations: \[i\hbar\partial_t\psi^{(s)} =
F_s\psi^{(s)}.\]
In contrast with \cite{GS:separation} we perform several
mathematical generalizations. By allowing more components in each $x_j$
(which we will indicate by using bold face) we introduce the possibility of internal
degrees of freedom such as spin, isospin, flavor, etc., and by considering species as a particular type of internal
degree of freedom, we suppress the species labels altogether. By this we achieve considerable notational
simplification along with ability to deal with multi-component
wave-functions. We also assume that the operators \(F\) can depend
explicitly on time. Such a generalization is necessary since certain
generators of infinitesimal symmetries, such as Galileian boosts, do
depend explicitly on time and it is necessary to treat these along with
\(F\) on equal mathematical footing. Furthermore, for theoretical
studies, one may want to ``switch on and off" certain terms by an
explicit time dependence. Our hierarchy
of equations will now be written as: \begin{equation}
\label{eq:hierarchy} i\hbar\partial_t\psi^{(n)} = F_n(t)\psi^{(n)}.
\end{equation} where the label $n$ indicates that we are dealing with
$n$ particles. We often drop this label, especially from the
wave-function, and sometimes for clarity we place it thus: \(F(t)_n\).
For the sake of mathematical generality, we will not always impose all
the requirements that physics may call for, maintaining the context at
what is mathematically natural for the level of abstraction adopted.

\begin{definition}
Given a set {\bf X},
called the {\em one-particle configuration space\/}, a {\em hierarchy of
multi-particle operators\/} is a family \(F\) of operators \(F_n\),
 \(n=1,2,\dots\), where \(F_n\) acts on a space of
functions \(\phi : {\bf X}^n \to {\bbf C}\) producing functions of the same
type.
By the {\em threshold\/} of a hierarchy we mean the
smallest integer \(c\) for which \(F_c \neq 0\).
\end{definition}
For our original context, the set \({\bf X}\) is \(S \times {\bbf R}^d\),
where \(S\) is a set of {\em species\/} and \({\bbf R}^d\) is the
configuration space for a single particle in a \(d\)-dimensional
Euclidean space. What makes \(S\) into a set of species as opposed to
some other type of internal degree of freedom is a specific assumption
about the form of the operator \(F_n\). Given \({\bf x} = ({\bf x}_1,\dots,{\bf x}_n)
\in {\bf X}^n\) let \({\bf x}_i = (s_i,x_i)\) with \(s_i \in S\) and \(x_i \in {\bbf
R}^d\). One can interpret a function \(\phi({\bf x})\) as a parameterized
family of functions \(\phi^{(s)}(x)\) considering the species labels as
parameters. If we now assume
\[(F_n\phi)({\bf x})= F_s(\phi^{(s)})(x)\]
where each \(F_s\) is some operator acting on functions defined on
\(({\bbf R}^d)^n\), then we have recovered the context of
\cite{GS:separation}.
When
\({\bf X}\) is a finite set we would
be dealing with quantum mechanics of a finite number of
degrees of freedom, and each evolution equations would be just a system
of ordinary differential equations in a finite dimensional space. In
this case the use of the word ``particle'' may be questionable, though
one could construe the equations as dealing with just the internal degrees of
freedom of particles, ignoring the spatial distribution.
Going in the other direction we could take
\({\bf X}=S\times {\bbf I}\times {\bbf R}^d\) where \(S\) is a set of species,
to be treated as explained above, and \({\bbf I}\) parameterizes the internal
degrees of freedom.  Presumably \({\bbf I}\) itself should be
conveniently written as a cartesian product \(\prod_\lambda {\bbf
I}_\lambda\) over the different types of internal degrees of freedom. We
would now be dealing with spatially distributed
particles of different species with any number of internal degrees of
freedom.
We shall also admit hierarchies of operators that
depend on some additional parameters (such as time), and also ones where
the multi-particle functions all depend on some fixed set of additional
variables (such as time). Besides hierarchies we also
treat just isolated   \(n\)-particle operators \(F_n\) for some values
of \(n\) without these being associated to a hierarchy. Depending on the
context, we shall
denote by a capital roman letters \(F,\, G,\, H, \dots\)  either an
individual  \(n\)-particle operator for fixed
\(n\), or a whole hierarchy of operators.
In what follows we
reserve the greek letter \(\psi\) for time dependent functions, that is,
those defined on \({\bbf R}\times {\bf X}^n\), and the greek letter \(\phi\)
for those defined on \({\bf X}^n\). We denote by \(\psi(t)\) the parameterized
function on \({\bf X}^n\) given by \(\psi(t)({\bf x}) = \psi(t,{\bf x})\). The right-hand
side of (\ref{eq:hierarchy}) should of course strictly speaking be
written as \(F_n(t)(\psi^{(n)}(t))\).

We shall impose one condition on \(n\)-particle operators
and hierarchies which reflects arbitrariness in labeling distinguishable
physical particles. If \(\pi\) is any permutation of \(\{1,\dots,n\}\)
then for any \(n\)-tuple \(w = (w_1,\dots,w_n)\) we define
\(\pi w = (w_{\pi(1)},\dots,w_{\pi(n)})\) and for any function \(\phi\)
on \({\bf X}^n\), we define \((\pi\phi)({\bf x}) = \phi(\pi {\bf x})\).
\begin{definition}
Let \(F\) be an \(n\)-particle
operators. We say this operator satisfies the {\em permutation property\/}
if, using the notation of the previous paragraph,  for each permutation
\(\pi\):
\[
F(\pi\phi) = \pi(F(\phi)).
\]
 We say a hierarchy satisfies the permutation property if each
\(n\)-particle operator does.
\end{definition}

From now on we always, and implicitly, assume the permutation property as it simplifies
some of the combinatorics and leads to no loss of generality for
any physical applications.

Our approach is {\em a priori\/}, disregarding mathematical questions of
domains and existence and uniqueness of solutions to the initial value
problem.
It is however useful at times to refer to the actual evolution, if it
exists and is unique,  and we
denote by \(E(t_2,t_1)\) the evolution operator from \(t_1\) to \(t_2\). That is,
\((E(t_2,t_1)\phi)({\bf x})=\psi(t_2)({\bf x})\) where \(\psi\) is the
unique solution of the initial value problem (\ref{eq:hierarchy}) with
\(\psi(t_1)({\bf x}) = \phi({\bf x})\). We of course have the group law for
\(E\):
\begin{eqnarray} 
E(t''', t'')\circ E(t'',t') &=& E(t''',t')  \label{eq:Ecomp}\\
E(t,t) &=& I. \label{eq:Eid}
\end{eqnarray}
From its definition, the evolution operator is easily shown to satisfy:
\begin{eqnarray}\label{eq:firstd}
\hbar\frac{\partial}{\partial t'} E(t',t) &=& \ibar F(t') \circ E(t',t) \\
\label{eq:second}
\hbar\frac{\partial}{\partial t} E(t',t) &=& - \Fd  E(t',t) \cdot \ibar F(t)
\end{eqnarray}
where \(\ibar = -i\). We note that in (\ref{eq:second}) the factor
\(\ibar\) cannot be moved to the front of the Fr\'echet derivative since
this operator is only real-linear and not necessarily linear. Formal
properties of the evolution operators are often useful heuristically
even if one has not established existence and uniqueness theorems for
the evolution equations.

We now review the definition of the separation property as introduced in
\cite{GS:separation}. Let \(H\) be a hierarchy of operators, and for
\(j=1,\dots,r\) let \(\phi_j\)  be functions on
\({\bf X}^{n_j}\). Let \(n = n_1 + n_2 +\cdots + n_r\). Adopt the same
notation for time-dependent functions \(\psi_j\) treating time as just a
parameter.
\begin{definition}\label{def:sephy}
We say a hierarchy \(H\) is a {\em separating hierarchy\/} if in the
notation of the previous paragraph:
\begin{equation}\label{eq:sephy}
H_{n_1}(\phi_1)\cdot H_{n_2}(\phi_2)\cdots H_{n_r}(\phi_r) =
H_n(\phi_1\cdot\phi_2\cdots\phi_r).
\end{equation}
\end{definition}

\begin{definition}
We say a hierarchy of evolution equations (\ref{eq:hierarchy}) satisfies
the {\em separations property\/} just in case using the notation of
the paragraph prior to Definition~\ref{def:sephy},  whenever the \(\psi_j\) are solutions of
(\ref{eq:hierarchy}) for particle numbers \(n_j\),  then
\(\psi_1\cdot\psi_2\cdots\psi_r\) is a solution of (\ref{eq:hierarchy})
for particle number \(n\). We shall also say that such
a hierarchy of equations is a {\em separating\/} hierarchy.
\end{definition}

Note that for a separating hierarchy of evolution equations, the
corresponding hierarchy of operators given by the right-hand side of
(\ref{eq:hierarchy}) is not necessarily (and in general will not be) a
separating hierarchy of operators. This abuse of language should not
cause confusion however.
For a separating hierarchy of evolution equation however, the
corresponding hierarchy of {\em evolution  operators}, if it exists, will
be a separating hierarchy.

To analyze
 (\ref{eq:sephy})
we substitute \(k_j\phi_j\) for \(\phi_j\) where the \(k_j\) are
complex numbers with \(\prod_{j=1}^rk_j = 1\). The right-hand side
does not change while the left-hand side becomes
\begin{equation}\label{eq:HScaled}
H_{n_1}(k_1\phi_1)\cdot H_{n_2}(k_2\phi_2)\cdots H_{n_r}(k_r\phi_r)
\end{equation}
which thus must be independent of the \(k_j\).
Suppose now that for some particle number \(n_0\), for some function
\(\phi_0\), and for some point \({\bf x}_0\), one has \(H_{n_0}(\phi_0)({\bf x}_0) \neq 0\).
Let now \(r=2\), \(n_1=n_2=n_0\)
and \(\phi_1=\phi_2=\phi_0\), then the invariance of
(\ref{eq:HScaled}) implies
\(H_{n_0}(k\phi_1)H_{n_0}(k^{-1}\phi_2) =
H_{n_0}(\phi_1)H_{n_0}(\phi_2)\).
Now the {\em variables\/} in \(\phi_1\) and \(\phi_2\) in this
equation are different, but we can substitute for both the same point
\({\bf x}_0\) and conclude that for all \(k\), \(H_{n_0}(k\phi)({\bf x}_0) \neq 0 \).
Applying again the invariance of (\ref{eq:HScaled}) to the case
 \(r=2, n_1 = n, n_2 = n_0\) with \(\phi_2 = \phi_0\)
one has
\[
H_n(k\phi_1)   = {H_{n_0}(\phi_2)({\bf x}_0) \over H_{n_0}(k^{-1}\phi_2)({\bf x}_0)}H_n(
\phi_1) = c(k)H_n(\phi_1).
\]
Thus unless all the operators in the hierarchy vanish identically,
\(H_n(k\phi)
= c(k)H_n(\phi)\) for some complex function \(c(k)\), and in particular \(c(1) = 1\).
Using
this in (\ref{eq:HScaled}) we see that \(\prod_{j=1}^n c(k_j)\) must be
independent of the \(k_j\). This is an exponentiated version of a
functional relation solved in \cite{GS:separation} and based on that
derivation we
 conclude that any locally integrable solution is of the form:
\[
c(k) = e^{a\ln|k| +ib\arg k}
\]
for some complex numbers \(a, b\). We shall discuss such
functions in section~\ref{sec:MixedPowers}.
\begin{definition}\label{def:MixPowHom}
Let \(a\) and \(b\) be complex numbers, an operator \(H\) satisfying
\begin{equation}\label{eq:MixPowHom}
H(k\phi)= e^{a\ln|k| +ib\arg k}H(\phi)
\end{equation}
will be called {\em mixed-power homogeneous\/} and the numbers \(a\) and
\(b\) will be called respectively the {\em first} and {\em second
exponential index} of \(H\). We call the property expressed by
(\ref{eq:MixPowHom}) {\em mixed-power homogeneity}. When \(a =1\) and
\(b=1\) we say the operator is {\em strictly homogeneous\/} and we call
the corresponding property {\em strict homogeneity\/}.
\end{definition}
 
\begin{theorem}
The operators of a separating hierarchy are mixed-power homogeneous
with the same exponential indices for all operators.
\end{theorem}

This theorem of course applies to the evolution operators of a separating
hierarchy of evolution equations (\ref{eq:hierarchy}) in which case
the exponential
indices of \(E(t',t)\) are in general functions of \(t'\) and
\(t\).

For a separating hierarchy of evolution equations (\ref{eq:hierarchy}) the corresponding
operators hierarchy \(F\) satisfies the infinitesimal versions of
(\ref{eq:sephy}) and (\ref{eq:MixPowHom}). These were derived in
\cite{GS:separation} and will be simply stated here.
\begin{definition}
We say of an operator \(F\) that it is {\em mixed-logarithmic homogeneous\/}
with {\em first\/} and {\em
second logarithmic index} \(p\) and \(q\) respectively if it satisfies
\begin{equation}\label{eq:ex-hom}
F(k\phi) = kF(\phi) + k(p\ln|k| + iq \arg k)\phi.
\end{equation}
We call the property expressed by the above equation {\em mixed-logarithmic
homogeneity}.
\end{definition}
When \(p=0\) and \(q=p\) the operator is strictly homogeneous.
\begin{definition}
We say of a hierarchy \(F\) that
it is a {\em tensor derivation\/} if it satisfies
\begin{equation}\label{eq:TenDer}
{F_{n_1}(\phi_1) \over \phi_1} + \cdots +
{F_{n_r}(\phi_r) \over \phi_r} = {F_n(\phi_1\cdots\phi_r)
\over \phi_1\cdots\phi_r}.
\end{equation}
 \end{definition}
One can interpret (\ref{eq:TenDer}) as Leibnitz's rule for the tensor
product. Another useful way of seeing this is to multiply  both sides of
(\ref{eq:TenDer}) by \(\phi_1\cdots\phi_r\):
\begin{equation}\label{eq:TenDertwo}
F_{n_1}(\phi_1)\cdot\phi_2\cdots\phi_r + \cdots +
\phi_1\cdots\phi_{r-1}\cdot F_{n_r}(\phi_r) =
F_n(\phi_1\cdots\phi_r).
\end{equation}

In \cite{GS:separation} the following two theorems are proved:

\begin{theorem}
Each operator in a tensor derivation is mixed-logarithmic homogeneous
with the same logarithmic indices.
\end{theorem}
\begin{theorem}
The operator hierarchy \(F(t)\) of a separating hierarchy of
evolution equations is a tensor
derivation for all \(t\).
\end{theorem}

The common logarithmic indices of all the evolution operators thus
constitute new universal physical constants with the dimension of
energy.

Being a tensor derivation is the infinitesimal version of
(\ref{eq:sephy}) and being mixed-logarithmic homogeneous is the
infinitesimal version of (\ref{eq:MixPowHom}).

We can derive the relationship between the logarithmic indices \(p(t)\)
and \(q(t)\) of the operators \(F(t)\) and the exponential indices \(a(t',t)\)
and \(b(t',t)\) of the evolution operators \(E(t',t)\). Applying \(E(t',t)\)
to \(k\phi\), using the mixed-power homogeneity, and Equations
(\ref{eq:Eid}--\ref{eq:firstd}), one arrives after
a short calculation at:
\begin{eqnarray}\label{eq:evola}
&i\hbar{\partial \over \partial t'}a(t',t) = p(t') \re a(t',t) + iq(t') \im
a(t',t)& \\ \label{eq:evolb}
&i\hbar{\partial \over \partial t'}b(t',t) = q(t') \re b(t',t) + ip(t') \im
b(t',t)& \\ \label{eq:initab}
&a(t,t)=1= b(t,t).&
\end{eqnarray}
Thus given \(p(t)\) and \(q(t)\) one can in principle solve the above linear
initial-value problem to uniquely determine \(a(t',t)\) and \(b(t',t)\).
Reciprocally Given \(a(t',t)\) and \(b(t',t)\) one finds:

\begin{eqnarray}\label{eq:pfroma}
p(t)= \left. i\hbar{\partial \over \partial t'}a(t',t)\right|_{t'=t}\\
\label{eq:qfromb}
q(t)= \left. i\hbar{\partial \over \partial t'}b(t',t)\right|_{t'=t.}
\end{eqnarray}

\section{Mixed powers}\label{sec:MixedPowers}

Due to their ubiquity, the form \(e^{a\ln|k| +ib\arg k}\) on the
right-hand side of (\ref{eq:MixPowHom}) and, changing letters,  its
logarithm \(p\ln|k| + iq\arg k\) on the right-hand side of
(\ref{eq:ex-hom}) deserve special attention.

\begin{definition}
Let \(z=re^{i\theta}\) be a
non-zero complex number in polar form and \((a,b)\) a pair of complex numbers. By the {\em
mixed \((a,b)\) power} of \(z\) we mean the number
\begin{equation} \label{eq:MixedPower}
z^{(a,b)} = r^ae^{ib\theta}
\end{equation}
which raises each factors in the polar decomposition to its
own power.
\end{definition}

Because \(\theta\)  in (\ref{eq:MixedPower})  is defined only modulo \(2\pi
i\) there is some ambiguity in defining the mixed power. It can be
uniquely defined in any domain in which \(\arg z\) is single-valued
by
\begin{equation} \label{eq:GoodPower}
z^{(a,b)} = e^{a\ln|z| + ib \arg z}.
\end{equation}
This makes the definition unique modulo the chosen branch of \(\arg\). If
we choose a branch that includes \(1\) in its interior with \(\arg 1 =
0\),
one can interpret
Equation~(\ref{eq:GoodPower}) as defining a germ at \(1\) of a continuous function
\(f(z)\) with \(f(1)= 1\). Such germs form a real algebra-like structure with the algebra
sum of \(f\) and \(g\) being the product \(fg\), the algebra product of \(f\)
and \(g\) being the composition \(f\circ g\) and the algebra scalar product of
\(r \in {\bbf R}\) with \(f\) being the power \(f^r\). This structure
satisfies most but not all of the axioms of a real algebra. What fails is
one of the distributive laws: \(f\circ gh \neq (f\circ g)(f\circ h)\), and
one of the scalar product laws: \(f\circ g^r \neq (f\circ g)^r\). Within this
structure the
mixed powers of \(z\) however form a  true real subalgebra and we have:
\begin{theorem}
\begin{eqnarray*}
z^{(a,b)}z^{(c,d)} = z^{(a+b,c+d)} \\
\left(z^{(c,d)}\right)^{(a,b)} = z^{(a,b)(c,d)}
\end{eqnarray*}
where 
\[
(a,b)(c,d)= (a\re c + ib \im c,\, b \re d +ia \im d).
\]
\end{theorem}

The proof is an easy verification using (\ref{eq:GoodPower}). We note in
particular that \(z^{(1,-1)} = \bar z\) so that the above algebra of mixed
powers contains the complex conjugation. The following are also useful
relations
\begin{eqnarray*}
(a,a)(c,d)= (ac,ad),\\
(a,-a)(c,d)= (a\bar c,- a \bar d).
\end{eqnarray*}
From this it is easy to compute the multiplication table for the
generators
\[
E=(1,1),\quad B=(1,-1),\quad I=(i,i),\quad J=(i,-i).
\]
Of these \(E\) is the multiplicative identity, \(B\) is the complex
conjugation, \(I\) is the ``exchange'' of the logarithm of the modulus
with minus the argument, and \(J\) is the same preceded by B. One has
Table~\ref{tab:GenProdLaw}.

\begin{table}[ht]
\centering
\begin{tabular}{c|ccc}
  & B & I & J \\  \hline
B & E & -J & -I \\
I & J & -E & -B \\
J & I & -B & E \\
\end{tabular} 
\caption{Product law} \label{tab:GenProdLaw}
\end{table}
In particular the set \(\{\pm E, \pm B, \pm I, \pm J \}\) forms a group.
The commutator Lie bracket \([(a,b), (c,d)] =
(a,b)(c,d) - (c,d)(a,b)\) for the last three generators (\(E\) commutes with
everything) is found to be:
\begin{eqnarray}{}
[B,I] &=& -2J \\{}
[I,J] &=& -2B \\{}
[J,B] &=& 2I.
\end{eqnarray}
These are  relations for the Lie algebra \({\goth sl}(2,{\bbf
R})\). This identification also follows from Theorem~\ref{thm:PowAction}
below.

\begin{theorem}\label{thm:PowAction}
The association of the real-linear
transformation
\begin{equation}\label{eq:PowAction}
z \mapsto (a,b)\cdot z = a\re z + ib\im z
\end{equation}
to the mixed power germ \(z^{(a,b)}\) is
an algebra isomorphism between the algebra of mixed powers and the
algebra of real-linear endomorphisms of \({\bbf C}\).
One has:
\begin{eqnarray*}
(a,b).((c,d)\cdot z) = ((a,b)(c,d))\cdot z \\
\ln z^{(a,b)} = (a,b)\cdot  \ln z  \\
(a,b)(c,d) = ((a,b)\cdot c,\,(b,a)\cdot d).
\end{eqnarray*}
Furthermore, in the ordered real basis \((1,i)\) of \({\bbf C}\), the
matrix of transformation (\ref{eq:PowAction}) is
\begin{equation}\label{eq:AMat}
\left(\begin{array}{cc}
\re a & -\im b \\
\im a & \re b
\end{array}
\right).
\end{equation}
\end{theorem}

The proof is utterly straightforward.
We also have:
\begin{theorem}
Let \(p(z,a,b)= z^{(a,b)}\) then
\begin{equation} \label{eq:DifPower}
\Fd p(z,a,b)\cdot (\zeta,\alpha,\beta) =
 z^{(a,b)}\left((\alpha,\beta)\cdot \ln z +(a,b)\cdot
 \left(\frac{\zeta}{z}\right)\right)
\end{equation}
which in particular implies that  for \(f(z) = z^{(a,b)}\) that the rank of \(\Fd f(z)\)
is two, unless \(\re a\bar b = 0\) in which case it is one, unless \((a,b)= (0,0)\)
\end{theorem}

The proof of (\ref{eq:DifPower}) is an easy verification using (\ref{eq:GoodPower}) once we
note that \(\ln |z| = \re \ln z\) and \(\arg z = \im \ln z\). The statement
about the rank follows now from (\ref{eq:AMat}).

\section{Symmetries: General considerations}

We begin discussing symmetries of a single \(n\)-particle
evolution equation:
\begin{equation}\label{eq:evolution}
i\hbar\partial_t \psi = F(t)\psi,
\end{equation}
going over some very well known ideas and results. Such a review is
nevertheless appropriate due to the non-linear context.

Informally, by a {\em symmetry} of (\ref{eq:evolution}) we mean an
operator \(V\) acting on functions \(\psi(t, {\bf x})\) defined on  \({\bbf
R}\times {\bf X}^n\) such that whenever \(\psi\) is a solution of
(\ref{eq:evolution}) then \(V\psi\) is also a solution. Symbolically:
\begin{equation}\label{eq:symmimp}
(i\hbar\partial_t -F(t))\psi = 0 \Rightarrow (i\hbar\partial_t
-F(t))V\psi = 0.
\end{equation}
The immediate difficulty with this is
that (\ref{eq:symmimp}) is not an operator equation for \(V\) and merely
states that \(V\) maps the solution set of (\ref{eq:evolution}) into
itself. For {\em a priori\/} studies no knowledge of the solution set
can be assumed and the usual recourse is to find some operator
equation of which (\ref{eq:symmimp}) is a consequence. This is
practically possible only after having assuming some structure for the
operator \(V\), calculating \(i\hbar\partial_t(V\psi)\), and in the
resulting expression substituting \(i\hbar\partial_t\psi\) by \(F(t)\psi\) to
arrive at a true operator equation relating \(V\) and \(F(t)\). This
also means that there is no general theory of symmetries, only various
particular theories relative to a given operator equation and certain
additional constraints.

If \(V\) and \(W\) are symmetries then it is clear that \(V\circ W\)
also is. Thus under composition the set of symmetries forms a
semi-group, and the set of invertible symmetries whose inverse is also a
symmetry,  a group. Whether the
set of symmetries obtained from a particular operator equation and set
of constraints is closed under composition or inversion is another
matter, though this is often the case and it is convenient that it be so.

We shall in this paper only consider symmetries of the form
\begin{equation}\label{eq:VoftT}
(V\psi)(t,{\bf x}) = \left(V(t)\psi(T(t))\right)({\bf x})
\end{equation}
where \(V(t)\) is an operator that acts
on functions on \({\bf X}^n\), and \(T:{\bbf R}\to{\bbf R}\) is some
diffeomorphism. The most common form for \(T\) is affine: \(T(t)= at +b\)
which includes such transformations as time translation, time inversion
and time dilation. One justification for assuming form (\ref{eq:VoftT})
is precisely to be able to handle space-time symmetries with such time
coordinate transformations.

Purely heuristically, such a form is not as restrictive as it may seem
for if one can uniquely solve the initial value problem for
(\ref{eq:evolution}) then any solution \(\psi\) can be constructed from
any of its time instant values \(\psi(t)\) by \(\psi(t')=
E(t',t)\psi(t)\). We can denote this by \(\psi = S(t)\psi(t)\) where
\(S(t)\) is an operator that transforms functions defined on \({\bf X}^n\) to
ones defined on \({\bbf R}\times {\bf X}^n\). One can now use the left-hand
side of (\ref{eq:VoftT})  to define \((V(t)\phi)({\bf x}) =
(VS(T(t))\phi)(t,{\bf x})\). Such an argument must however be used with
caution if one is trying to avoid assuming any knowledge of the
evolution operator  or, what amounts to the same thing, the solution
set.

Our basic form for symmetries is invariant under composition and
inversion and we easily show:
\begin{theorem}\label{thm:symmcomp}
If \(V\) and \(W\) are both of the form
(\ref{eq:VoftT}) then so is \(V\circ W\) where we have:
\begin{eqnarray*}
(V\circ W) (t) &=& V(t)\circ W(T_V(t))\\
T_{V\circ W} &=& T_W\circ T_V.
\end{eqnarray*}
Furthermore, if \(V\) is invertible, then \(V^{-1}\) is also of the form
(\ref{eq:VoftT}) and we have:
\begin{eqnarray*}
V^{-1}(t) &=& V(T_V^{-1}(t))^{-1}\\
T_{V^{-1}} &=& T_V^{-1}.
\end{eqnarray*}
\end{theorem}

In term of the evolution operator \(E(t',t)\) for (\ref{eq:evolution}),
the property of \(V\) being a symmetry is now expressed through
\begin{equation}\label{eq:ESymm}
V(t')\circ E(T(t'),T(t)) = E(t',t) \circ V(t).
\end{equation}
From this one determines by the group law for \(E\) that
\begin{equation}\label{eq:symmisE}
V(t) = E(t,0)\circ V(0)\circ E(T(0),T(t)).
\end{equation}
Thus any symmetry has the form (\ref{eq:symmisE}) where \(V(0)\) is
arbitrary (as long as it transforms proper initial data into proper
initial data). While this is undoubtedly true, it trivially reduces all
symmetries to the knowledge of the evolution operator which for
practical purposes is quite unproductive. This is yet one more
indication that one generally only obtains
useful information from symmetries if these belong to a restricted class
of operators.

To obtain the operator equation for a symmetry we
differentiating \(V(t)\psi(t)\) with respect to \(t\) and use
(\ref{eq:evolution}) for both \(\psi\) and \(V\psi\) to deduce:
\begin{equation}\label{eq:evolsymm}
\hbar\frac{\partial V(t)}{\partial t} =\ibar F(t)\circ V(t) - T'(t){\bbf
D}V(t)\cdot \ibar F(T(t)).
\end{equation}
Under general conditions on the solubility and uniqueness of the
the initial-value
problem, equation (\ref{eq:evolsymm}) is necessary and sufficient for
\(V\) to be a symmetry. It is thus an
appropriate formal definition of symmetry for {\em a priori\/} considerations.

\begin{definition}
We say an operator \(V\) given by (\ref{eq:VoftT}) is a (formal) {\em symmetry\/} of the evolution
equation (\ref{eq:evolution}) if condition (\ref{eq:evolsymm}) holds.
\end{definition}

For \(T(t)= t\) and real-linear operators, the right-hand side of (\ref{eq:evolsymm})
would be a usual commutator and this would be a familiar condition. If
\(T'(t) \neq 1\) then even for
real-linear operators and \(F(t)\) time independent, the right-hand side
in general would be a deformed or ``quantum" commutator. We
don't explore  condition (\ref{eq:evolsymm}) in its general form in any
detail.

Solutions of (\ref{eq:evolsymm}) are subject to the operations of
Theorem~\ref{thm:symmcomp}:
\begin{theorem}
If both \(V\) and \(W\) satisfy their respective equations
(\ref{eq:evolsymm}), then so does \(V\circ W\). Furthermore if \(V\) is
invertible, then \(V^{-1}\) also satisfies its equation
(\ref{eq:evolsymm}).
\end{theorem}
Note that this theorem does not say we can in general compose or invert
solutions to  equation (\ref{eq:evolsymm}) for a {\em fixed\/} function \(T\)
since this function generally changes  when we compose or invert.

A significant simplification occurs in our theory of symmetries when we
deal with a
one-parameter group of symmetries, that is, symmetries \(V(r), r \in
{\bbf R}\) such that
\(V( r+s) = V( r)\circ V( s)\) and \(V 0) = I\). One would
have:
\begin{equation}
(V(r)\psi)(t,{\bf x}) = \left(V(t,r)(\psi(T(t,r))\right)({\bf x}).
\end{equation}
If we now write
\(V( r) = I + rK + o(r)\) and similarly \(V(t) = I + rK(t)+ o(r)\), and
\(T(t)=t+r\tau(t)+ o(r)\), then one has for \(K\), the {\em
infinitesimal generator \/} of \(V(r)\), the following form:
\begin{equation}\label{eq:InfGen}
(K\psi)(t,{\bf x}) = (K(t)\psi(t))({\bf x}) +
\tau(t)(\partial_t\psi)(t,{\bf x}).
\end{equation}
If now in (\ref{eq:evolsymm}) we  evaluate the derivative with
respect to \(r\) at \(r=0\) we get:
\begin{equation}\label{eq:symmbrak}
\hbar\frac{\partial K(t)}{\partial t} =
 [\ibar F(t), K(t)] - \frac{\partial}{\partial t}(\tau(t)\ibar F(t)).
\end{equation}
The first term on the right-hand side is of course the bracket of the two operators
considered as vector fields. The second term is the remnant of the
deformed nature of the ``bracket'' on the right-hand side of
(\ref{eq:evolsymm}):
\begin{definition}
We say an operator \(K\) having the form (\ref{eq:InfGen}) is
an  {\em  infinitesimal symmetry\/} of (\ref{eq:evolution}) if and only
if (\ref{eq:symmbrak})
holds.
\end{definition}

For real-linear operators the bracket reduces to the usual commutator.  For
non-linear theories we've seen that for two operators \(F, G\) there are
in fact three notions of commutator that generalize the usual
one. The first would be the ``true" commutator  \(F\circ G -
G\circ F\), the second can be obtained from it by replacing \(F\) by
\(I+rF\) and evaluating the derivative with respect to \(r\) at \(r=0\),
which gives us \(F\circ G - \Fd G\cdot F\), and the
third can be obtained by
now replacing \(G\) with \(I+rG\) and again evaluating the derivative with
respect to \(r\) at \(r=0\), which gives \([F,G]\). Of these only the third is a
Lie bracket, and the second is not even anti-symmetric. Each one is the
appropriate generalization of the real-linear commutator in the right
context. One also has the ``group" commutator \(F\circ G \circ
F^{-1}\circ G^{-1}\). This
in fact is
related to the ``true'' commutator through \(F\circ G \circ F^{-1}\circ G^{-1} = I
+ F\circ (G \circ F^{-1} -F^{-1}\circ G)\circ G^{-1}\) and so no
essentially new forms are introduced by this construct.
One can also get
deformed or ``quantum'' versions by starting with a deformation of the
true commutator and then proceeding in the same manner, allowing also
for the deformation to be subject to an expansion in a small parameter
\(r\).

The composition property of symmetries given by Theorem~\ref{thm:symmcomp}
of course has its counterpart for
infinitesimal symmetries:
\begin{theorem}
If \(K\) and \(L\) are two infinitesimal
symmetries of (\ref{eq:evolution}) then so is \([K, L]\).
Furthermore,
\begin{eqnarray}\label{eq:KbraL}
[K,L](t)&=&[K(t),L(t)] + \tau_K(t)\frac{\partial L(t)}{\partial t}
-\tau_L(t)\frac{\partial K(t)}{\partial t}\\ \label{eq:taubrak}
\tau_{[K,L]}(t)&=&\tau_K(t)\tau'_L(t) - \tau_L(t)\tau_K'(t).
\end{eqnarray}
\end{theorem}
The proof of (\ref{eq:KbraL}) is a straightforward  calculation. To show
that \([K,L]\) again satisfies (\ref{eq:symmbrak}) is an exercise in the
Jacobi identity. The infinitesimal version of inversion is negation.
Obviously if \(K\) satisfies (\ref{eq:symmbrak}) then so does \(-K\).
The right-hand side of (\ref{eq:taubrak}) is the Lie bracket of
\(\tau_K\) and \(\tau_L\) considered as vector fields on \({\bbf R}\).

One can now pose two problems, one inverse to the other: for a given
\(F(t)\) find all operators of our type \(V\) or \(K\),
within some convenient class, that satisfy respectively (\ref{eq:evolsymm}) or
(\ref{eq:symmbrak}), or inversely, given a set of such operators
determine the \(F(t)\), within some convenient class, that have these
transformations as symmetries. Such problems are often tractable but
require considerable calculations and are now generally attacked by
algebraic computation systems.  See for example \cite{Champagne} and
references therein. We shall not pursue these question in any detail.

From a purely general perspective Equations~(\ref{eq:evolsymm}) and
(\ref{eq:symmbrak}) are
not very enlightening. They are evolution equations for \(V(t)\) and
\(K(t)\) respectively and given arbitrary \(V(0)\) or \(K(0)\) could in
principle be solved to provide \(V(t)\) or \(K(t)\). This is a reflection
of the trivialization of symmetries to the evolution operator that we've
mentioned earlier.
Equations~(\ref{eq:evolsymm}) and (\ref{eq:symmbrak}) are always
supplemented by additional conditions.
Typically these are
that \(V(t)\) or \(K(t)\) be independent of time or that they depends on
time in a specified manner, that they be differential operators of a
specified type, that they reflect space-time transformations, etc. Such
additional restrictions generally transform equations
(\ref{eq:evolsymm}) and (\ref{eq:symmbrak}) into overdetermined systems.

\section{Separating symmetries}
One restriction that is quite natural for some symmetries
that reflect invariance under an action that can be performed by the
experimenter (such as change of inertial frame) is once again that
systems consisting of uncorrelated parts continue being so after the
symmetry transformation, and each part transforms as if the other parts
did not exist. This is a direct extension of the separation property to
symmetries. Of course we now deal with a hierarchy of symmetries \(V_n\)
or of infinitesimal symmetries \(K_n\) for each particle number \(n\).
One must be careful drawing conclusion from the separation property for
a symmetry since the hierarchy of symmetries is composed of operators
that act on function on \({\bbf R}\times {\bf X}^n\) and not on \({\bf X}^n\).
In particular, a separating
symmetry is not a mixed-power homogeneous operator since solutions to
(\ref{eq:evolution}) do not scale if \(F(t)\) is not strictly
homogeneous. As was shown in \cite{GS:separation}, if \(F(t)\) is
mixed-logarithmic homogeneous, then solutions scale by  time-dependent
factors \(w(t)\) comprised of  the solution of the equation
\(i\hbar\partial_t\ln w(t) = (p(t),q(t))\cdot \ln w(t)\). A symmetry
will have homogeneity properties only with respect to such
time-dependent multipliers. If we write down
however what the separation property means for a symmetry of the form
(\ref{eq:VoftT}), we immediately find that this implies that \(V(t)\) must be a
separating hierarchy.

\begin{definition}\label{def:hierarchy}
Given a hierarchy of evolution equations (\ref{eq:hierarchy})
we say a hierarchy \(V\), respectively \(K\), of operators  is
a {\em symmetry \/}, respectivley {\em infinitesimal symmetry\/}, of the hierarchy of
equations if each \(V_n\)  is a symmetry, respectively each \(K_n\) is an
infinitesimal symmetry,
of the corresponding equation.
Given a separating hierarchy of evolution equations we say of a hierarchy of
symmetries of form (\ref{eq:VoftT}), respectively (\ref{eq:InfGen}), that it
satisfies the {\em separation property}, or that it is a {\em separating
symmetry\/} if,  for all \(t\), \(V(t)\) is a separating hierarchy, respectively
\(K(t)\) is a tensor derivation.
\end{definition}

One then concludes that a for a separating symmetry the \(V_n(t)\) are
mixed-power homogeneous with the same exponential indices
and that for an infinitesimal separating symmetry the
\(K_n(t)\) are  mixed-logarithmic homogeneous with the same logarithmic
indices.

The homogeneity indices of symmetries and infinitesimal symmetries are
related to those of the operators in the evolution equations through
their own evolution equations. We have:
\begin{theorem}
Let \(V\) of form (\ref{eq:VoftT}) be a symmetry of a separating hierarchy of evolution equations
and let \((a(t),b(t))\) be the exponential indices of \(V(t)\) and
\((p(t),q(t))\) be the logarithmic indices of \(F(t)\). One has
\begin{eqnarray}\label{eq:evolInd}
\lefteqn{\hbar\frac{d}{dt}(a(t),b(t)) = } \nonumber \\
& & (\ibar p(t),\ibar q(t))\cdot (a(t),b(t)) -
T'(t)(a(t),b(t))\cdot(\ibar p(T(t)),\ibar q(T(t))).
\end{eqnarray}
Furthermore if now \(K\) of form (\ref{eq:InfGen}) is an infinitesimal
symmetry of the same hierarchy of evolution equations, and if now
\((c(t),d(t))\) are the logarithmic indices of \(K(t)\), then
\begin{equation}\label{eq:evolind}
\hbar\frac{d}{dt}(c(t),d(t)) = [(\ibar p(t),\ibar q(t)),(c(t),d(t))]
-\frac{d}{dt}(\tau(t)(\ibar p(t),\ibar q(t))).
\end{equation}
Finally if \(K\) is the infinitesimal generator of the one-parameter
group \(V(r)\), then
\[
(a(t,r),b(t,r)) = (1,1)+r(c(t),d(t))+o(r).
\]
\end{theorem}
The theorem is easily proved by applying (\ref{eq:evolsymm}) and
(\ref{eq:symmbrak}) to \(k\phi\) and using the homogeneity properties of
the operators involved. As a short-cut for (\ref{eq:evolInd}) one can
apply (\ref{eq:ESymm}) to \(k\phi\) and use
(\ref{eq:evola}--\ref{eq:qfromb}). This of course is only legitimate if
the evolution operators exist, but the formal result is true in any
case. Likewise (\ref{eq:evolind}) follows
directly from (\ref{eq:symmbrak}) and Theorem~\ref{thm:DerBrak} below.

We shall from now on deal only with separating symmetries.

\section{Lie algebra and liftings of tensor derivations}\label{sec:Laf}

We first prove an analog of the well known Euler's equation for
homogeneous functions.

\begin{theorem}\label{thm:Euler}
Let \(H\) be a mixed-power homogeneous operator with exponential indices \((a,b)\),
\(K\) be a mixed-logarithmic homogeneous operator with logarithmic indices
\((p,q)\), and \(\eta\) any complex number, then
\begin{eqnarray}\label{eq:Eulerpow}
{\bbf D}H(\phi)\cdot \eta\phi &=& (a,b)\cdot\eta H(\phi) \\
\label{eq:Eulerlog}
\Fd K(\phi)\cdot \eta\phi &=&
 \eta K(\phi)+ (p.q)\cdot\eta\, \phi.
\end{eqnarray}
\end{theorem}
{\em Proof:\/} One has
\begin{eqnarray*}
H((1+r\eta)\phi)&=& (1+r\eta)^{(a,b)}H(\phi), \\
K((1+r\eta)\phi)&=& (1+r\eta)K(\phi)+
(1+r\eta)\ln(1+r\eta)^{(p,q)}\phi.
\end{eqnarray*}
Using (\ref{eq:DifPower}) one can
evaluate the derivative of these
with respect to \(r\) at \(r=0\), and obtain (\ref{eq:Eulerpow}) and
(\ref{eq:Eulerlog}) respectively.
Q.E.D.

Given two infinitesimal generators
\(F\) and
\(G\) of one-parameter groups \(V(r)\) and \(W(r)\) one has \[
\lim_{n\to\infty}(V(r/n)W(r/n)V(-r/n)W(-r/n))^{n^2} = I +r^2[F, G]
+o(r^4).\] Thus, modulo the possibility of exponentiation, one can show
that the Lie bracket preserves those properties of operators that
have expression in the corresponding
exponentiated groups and there behave appropriately under compositions.
In particular it is not surprising that tensor derivations form a Lie
algebra:
\begin{theorem}\label{thm:DerBrak}
\hfill
\parskip=0pt
\begin{enumerate}
\item If \(F\) and \(G\) are mixed-logarithmic
homogeneous operators with indices \((p_F,q_F)\) and \((p_G,q_G)\) respectively,
then so is
\([F,G]\) with logarithmic indices
\begin{equation}\label{eq:DerIndBrak}
(p_{[F,G]},q_{[F,G]}) = [(p_F,q_F),(p_G,q_G)].
\end{equation} 
\item If \(F\) and \(G\) are tensor derivations then so
is \([F,G]\) with logarithmic indices given by (\ref{eq:DerIndBrak}).
The threshold of \([F,G]\) is greater than or equal to the maximum of the
thresholds of \(F\) and \(G\).
\end{enumerate}
\end{theorem}

{\em Proof:\/}
To prove the first part we first note that:
\[\Fd F(k\phi)\cdot G(k\phi) = \left.{d \over ds} F(k\phi +
s G(k\phi))\right|_{s=0.}\] The term to be differentiated above is equal
to:
\[ F(k\phi + ks(G\phi + \ln k^{(p_G,
q_G)}\phi)) =\]
\[k( F(\phi+s G\phi + s\ln k^{(p_G, q_G)}\phi) +  \ln
k^{(p_F,q_F)}(\phi +s G\phi + s\ln k^{(p_F,
q_F)}\phi)).\]
The derivative of this with respect to \(s\) at \(s=0\) is:
\[k\left(\Fd F(\phi)\cdot G(\phi) + \Fd F(\phi)\cdot \ln
k^{(p_G,q_G)}\phi + \ln k^{(p_F,q_F)}(G(\phi) +
 \ln k^{(p_G,q_G)}\phi)\right).\]
From this,
\begin{eqnarray*}
\lefteqn{[F,G](k\phi) = k[F,G](\phi) + \mbox{}}\\
& & \mbox{} + k(\Fd F(\phi)\cdot \ln
k^{(p_G,q_G)}\phi - \ln k^{(p_G,q_G)}F(\phi)) + \mbox{}  \\
& & \mbox{}-k(\Fd G(\phi)\cdot \ln k^{(p_F,q_F)}\phi - \ln
k^{(p_F,q_F)}G(\phi)).
\end{eqnarray*}
 Using the generalized Euler's formula (\ref{eq:Eulerlog}) in this
expression we deduce the
formula for \((p_{[F,G]}, q_{[F,G]}) \) and prove the first part.

For the case of tensor derivations, one has, using the notation of Definition~\ref{def:sephy}:
\[\Fd F_n(\phi_1\cdots\phi_r)\cdot G_n(\phi_1\cdots\phi_r) =\]
\[{d \over ds} F_n(\phi_1\cdots\phi_r +
s G_n(\phi_1\cdots\phi_r))|_{s = 0} =\]
\[{d \over ds}(F_n(\phi_1\cdots\phi_r +s
G_{n_1}(\phi_1)\cdot\phi_2\cdots\phi_r +
\cdots + s \phi_1\cdot\phi_2\cdots
\phi_{r-1}\cdot G_{n_r}(\phi_r))|_{s=0.} \]
The quantity being differentiated differs by a term of order
\(o(s^2)\) from
\[F_n((\phi_1 + s G_{n_1}\phi_1)\cdots(\phi_r +
s G_{n_r}\phi_r)).\]
Using the fact that \(F\) is a tensor derivation we can apply to this
expression Leibnitz's
rule (\ref{eq:TenDertwo}), evaluate the
derivative with respect to \(s\) at \(0\), and arrive at:
\begin{equation}\label{eq:FpG}
\sum_{j=1}^r \Fd F_{n_j}(\phi_j)\cdot G_{n_j}(\phi_j)\cdot\hat
\phi_j + \sum_{j\neq k} F_{n_j}(\phi_j)\cdot
G_{n_k}(\phi_k)\cdot \hat\phi_{jk}
\end{equation}
where we've introduced the partial (tensor)
products \(\hat\phi_j = \prod_{i\neq j}\phi_i\) and \(\hat\phi_{jk} =
\prod_{i\neq j,k}\phi_i\), and where all the tensor
products in (\ref{eq:FpG}) are
to be interpreted as occurring in the original order of
\(\phi_1\cdots\phi_r\).  To not be misled by the notation in
(\ref{eq:FpG}), we mention
that the first in-line dot in the first term designates an application of
a Fr\'echet derivative to a vector, while in the second term it
designates
a tensor product. As the second term in (\ref{eq:FpG})
is symmetric under the interchange of \(F\) and \(G\), the
corresponding  terms in \([F,G](\phi_1\cdots\phi_r)\) cancel and we
deduce that the bracket satisfies the separation property.  The
statement about the thresholds is obvious. Q.E.D.

We now review the lifting properties of tensor
derivations.

Let \(F\) be an operator acting on functions from \({\bf X}^n\) to
\({\bbf C}\) producing functions of the same type. Let now \(m>n\) and
\(J = (j_1,\dots,j_n)\) be an \(n\)-tuple of distinct elements of
\(\{1,\dots,m\}\) in increasing order. A function
\(\phi({\bf x}_1,\dots,{\bf x}_m)\) can be construed as a parameterized family of
functions \( \phi_{\bf y}({\bf x}_{j_1},\dots,{\bf x}_{j_n})\) by taking each \({\bf x}_k\) for
\(k \not\in \{j_1,\dots,j_{n_j}\}\) as a parameter \({\bf y}_k\). Applying
\(F\) to each member of this family one gets another parameterized
family of functions \(F(\phi_{\bf y})({\bf x}_{j_1},\dots,{\bf x}_{j_n})\) which we can
reinterpret back as a function \(F^J(\phi)({\bf x}_1,\dots,{\bf x}_m)\). This
defines a new operator \(F^J\).
\begin{definition}\label{def:lifting} The operator \(F^J\) defined in the
previous paragraph is called a {\em lifting} of \(F\).
\end{definition}

The three following theorems were proved in \cite{GS:separation}.

\begin{theorem}\label{thm:Fonelift}
Let \(F\) be a one-particle mixed-logarithmic homogeneous
operator with logarithmic indices \(p\) and \(q\). For \(n
\geq 1\) define  \(n\) particle operators by
\begin{equation}\label{eq:Fonelift}
F^\#_n\phi = \sum^n_{j=1}F^{(j)}\phi -(n-1) (p, q)\cdot
\ln\phi\,\phi.
\end{equation}
The resulting hierarchy  \(F^\#\) is a tensor derivation extending \(F\)
(called the {\em canonical lifting\/} of \(F\)).
\end{theorem}
\begin{theorem}\label{thm:Felllift}
Let \(\ell > 1\) and \(F\) be a strictly homogeneous \(\ell\)-particle operator
which vanishes on any tensor product
function. For \(n\geq \ell\) define
\(n\) particle operators by
\begin{equation}\label{eq:Felllift}
F^\#_n = \sum_J F^J
\end{equation}
where the sum runs over all \(J=(j_1,\dots,j_\ell)\) of \(\ell\)-tuples of
distinct elements of \(\{1,\dots,n\}\) in increasing order.
The resulting hierarchy  \(F^\#\)
is a tensor derivation of threshold \(\ell\) extending \(F\) (called the {\em canonical lifting\/} of \(F\)).
\end{theorem}
One sees that in (\ref{eq:Fonelift}) if \(p\) and \(q\) vanish
then (\ref{eq:Fonelift}) can be construed as the \(\ell = 1\) case of
(\ref{eq:Felllift}). It is sometimes useful, in spite of the
breach of good notational discipline, to write a single
formula:
\begin{equation}\label{eq:badbad}
F^\#_n\phi = \sum_J F^J\phi -(n-1)(p, q)\cdot \ln\phi\,\phi
\end{equation}  to cover
both cases in a single argument with the understanding that the second
term is zero for \(\ell \neq 1\).

\begin{theorem}\label{thm:CanDecomp}
 Let  \(F\) be a tensor derivation. Define
derivations \(d_jF\) as follows: \[d_1F = F_1{}^\#,\] and having defined
\(d_1F,\dots,d_rF\), let \[d_{r+1}F = (F - {\textstyle \sum_{j=1}^r}
d_jF)_{r+1}{}^\#.\]
One has \(F =
\sum_{j=1}^\infty d_jF\)  (called the {\em canonical decomposition
of \(F\)\/}), the \(d_j\) are real-linear idempotents, and if
\(d_jF\) is not zero, its threshold is \(j\).
Conversely if for each \(j\) we are given a \(j\)-particle operator
\(F_{(j)}\) satisfying:
\begin{enumerate}
\item \label{item:onegen}\(F_{(1)}\) is
mixed-logarithmic homogeneous.
\item \label{item:highgen}For \(j
> 1\), \(F_{(j)}\) is strictly homogeneous and vanishes on
tensor product functions;
\end{enumerate}
then the derivation \(F = \sum_{j=1}^\infty F_{(j)}{}^\#\)
satisfies \(d_jF = F_{(j)}{}^\#\).
\end{theorem}

These theorems provide us with a canonical procedure to construct tensor
hierarchies by the introduction  of new generators
at each particle number threshold. The operator \((d_jF)_j\) is called
the {\em canonical generator\/} of \(F\) at threshold \(j\). These
uniquely define the hierarchy and are themselves objects that can be
freely given subject only to conditions (\ref{item:onegen}) and
(\ref{item:highgen}) above. In physical theories, generators at particle
numbers greater than one introduce truly new effects in correlated
systems that are absent for smaller number of particles.

\begin{definition}
We call a  \(j\)-particle operator \(F\) a {\em
generator\/} if it satisfies item (\ref{item:onegen}) or
(\ref{item:highgen}) of Theorem~\ref{thm:CanDecomp}.
\end{definition}

The lifting properties of the Lie bracket are quite complex. Suppose we
are given two one-particle generators \(F\) and \(G\). Let
\(H = [F,G]\) define a third one,  and let
\(F^\#\), \(G^\#\), and \(H^\#\) be the corresponding
canonical liftings. It is not generally true that \(H^\# =
[F^\#, G^\#]\). This is a purely non-linear effect and has the
consequence that if one has a set of one-particle symmetries  of a
one-particle evolution equation then the canonically lifted
multi-particle equations are not necessarily symmetric under the
canonically lifted one-particle symmetries. Since one naively would
expect one-particle symmetries to be extensible to multi-particle
symmetries, especially if no new multi-particle effects are introduced
through new canonical generators, this question bears examining.

\begin{definition}
For a pair of complex numbers \((a,b)\) define the \(n\)-particle
operator \(\Lambda(a,b)\) by
\[
\Lambda(a,b)\phi = (a,b)\cdot\ln\phi \, \phi.
\]
For a mixed-logarithmic homogeneous operator \(F\) with indices \((a,b)\)
let \(\Lambda_F = \Lambda(a,b)\). Define then \(F^{\natural}\) by:
\begin{equation}\label{eq:natural}
F=F^\natural + \Lambda_F.
\end{equation}
\end{definition}

One easily verifies:
\begin{theorem}\label{thm:counbrak}
\hfill
\parskip=0pt
\begin{enumerate}
\item \(\Lambda(a,b)\) is mixed-logarithmic homogeneous
with logarithmic indices
\((a,b)\).
\parskip=\skipu
\item \label{item:counbrak}\([\Lambda(a,b),\Lambda(c,d)] = \Lambda([(a,b),(c,d)])\).
\item The canonical lifting of the one-particle \(\Lambda(a,b)\) to an
\(n\)-particle operator is the corresponding \(n\)-particle \(\Lambda(a,b)\).
\item  \label{item:brakrep} If \(F\) and \(G\) are  mixed-logarithmic
homogeneous operators, then \(\Lambda_{[F,G]} = [\Lambda_F,\Lambda_G]\).
\item For a one-particle generator \(F\)
with logarithmic indices \((a,b)\),
the canonical lifting \(F^\#\)satisfies:
\[
F^\#_n = F_n^{\natural \#} + \Lambda(a,b).
\]
\end{enumerate}
\end{theorem}

Item~(\ref{item:counbrak}) shows that \((a,b) \to \Lambda(a,b)\) is a
representation of the mixed-power Lie algebra.

\begin{theorem}\label{thm:obsatl}
\hfill
\parskip=0pt
\begin{enumerate}
\item Let \(F\) be an \(\ell\)-particle generator and \(G\) an
\(m\)-particle generator with \(1 \leq \ell \leq m\).
Let \(F^\#\) and \(G^\#\) be their respective canonical liftings.
For any particle number \(n\) with \(n > m\):
\begin{equation}\label{eq:liftdeltal}
[F^\#_n, G^\#_n]-[F^\#_m,G]^\#_n = \sum_K \sum_{J \not\subset K}
[F^{\natural J},G^{\natural K}]
\end{equation}
where \(J\) is an \(\ell\)-tuples \((j_1,\dots,j_\ell)\) of
elements of \(\{1,\dots,n\}\) in increasing order, \(K\) is an
\(m\)-tuple of the same type and where we write \(J \not\subset K\) to
mean \(\{j_1,\dots,j_\ell\} \not\subset \{k_1,\dots,k_m\}\).
\item The obstruction to the equality
\begin{equation}\label{eq:comlifbrakl}
[F^\#,G^\#]= [F^\#_m,G]^\#
\end{equation}
is the set of operators on the right hand side of
(\ref{eq:liftdeltal}) at particle numbers from \(m+1\) to \(m+\ell\).
These operators are zero if and only if (\ref{eq:comlifbrakl}) holds.
\end{enumerate}
\end{theorem}

{\em Proof:\/} We first note that \([F^\#_m,G]=  [F^\#,
G^\#]_m\) so the left-hand side is a legitimate
\(m\)-particle generator (even though \(F^\#_m\) itself is
not) since by Theorems \ref{thm:DerBrak}, \ref{thm:Fonelift} and
\ref{thm:Felllift},  \([F^\#,G^\#]\) is a tensor
derivation of threshold at least \(m\).

To expedite the proof we use the notational shortcut indicated by
(\ref{eq:badbad}) and explained in that paragraph. We shall also employ
set-theoretic notation such as ``\(J \subset K\).'' This
is to be understood as referring to the underlying sets.

One has for \(n\)-particle operators:
\[[F^\#_n, G^\#_n] = [\sum_JF^J -(n-1)\Lambda_F,
\sum_KG^K -(n-1)\Lambda_G]\]
which is
\begin{eqnarray*}
\lefteqn{\sum_K\sum_{J \subset K} [F^J,G^K ]
+\sum_K\sum_{J \not \subset K} [F^J,
G{(k)}]+\mbox{}} \\
& & \mbox{}-(n-1)\sum_J[F^J, \Lambda_G]
-(n-1)\sum_K[ \Lambda_F,G^K] +
(n-1)^2[ \Lambda_F,
 \Lambda_G].
\end{eqnarray*}
Now
\[[F^\#_m,G]^\#_n = \sum_K\sum_{J \subset K} [F^J,G^K ] -
(n-1)\Lambda_{[F,G]}.\]
So, using Item~\ref{item:brakrep} of
Theorem~\ref{thm:counbrak}, \([F^\#_n, G^\#_n]-[F^\#_m,G]^\#_n\)
is found to be:
\begin{eqnarray*}
\lefteqn{\sum_K\sum_{J \not \subset K} [F^J,
G^K] + \mbox{}}\\
& & -(n-1)\sum_J[F^J, \Lambda_G]
-(n-1)\sum_K[ \Lambda_F,G^K] +
n(n-1)[ \Lambda_F,
 \Lambda_G].
\end{eqnarray*}
Substituting into this expression \(F^J =
F^{\natural J}+ \Lambda_F\) and \(G^K =
G^{\natural K}+ \Lambda_G\) one arrives after a short calculation
at (\ref{eq:liftdeltal}). To deduce the statement about the obstruction
let \(n > m + \ell \) and consider the right-hand side of
(\ref{eq:liftdeltal}). Let \(\Gamma\) be the set of all pairs \((J,K)\)
of \(m\) and \(\ell\)-tuples of elements of \(\{1,\dots,n\}\) in
increasing order with \(J \not \subset K\). For \(\gamma =(J,K)\in \Gamma\) let
\(C(\gamma) =[F^J,G^K ] \). For any subset \(\Omega \) of
\(\Gamma\) let \(C(\Omega)= \sum_{\gamma \in \Omega}C(\gamma)\). The
right-hand side of (\ref{eq:liftdeltal}) is \(C(\Gamma)\). Let
\(I_1,\dots,I_d\) be the enumerated distinct \((m+\ell)\)-tuples of elements of
\(\{1,\dots,n\}\) in increasing order. Now, since the number of
variables affected by any \(C(\gamma)\) is at most \(m+\ell\), one has
 \(\Gamma = \bigcup_{i=1}^d \Gamma_i\) where \(\Gamma_i\) consists of those pairs \((J,K)\) for
which \(J \subset I_i\) and \(K \subset I_i\). One has the classic
formula:\[C(\Gamma) = \sum_{k=1}^d (-1)^{k+1} \sum_{i_1<i_2<\cdots <
i_k}C(\Gamma_{i_1}\cap\Gamma_{i_2}\cap\cdots\cap\Gamma_{i_k}).
\]
Consider now the term
\(C = C(\Gamma_{i_1}\cap\Gamma_{i_2}\cap\cdots\cap\Gamma_{i_k})\). This is a
sum over all pairs \((J,K)\)  for which \(J,K \subset I = I_{i_1}\cap
I_{i_2}\cap\cdots
\cap I_{i_k}\). Suppose the set of such pairs is not empty, so that the
number of elements in \(I\) is some number \(h\) with \(m+1 \leq h \leq
m+\ell\). When \(C\) is applied to a function \(\phi\) on \({\bf X}^n\), all the
variables not indexed by an element of \(I\) can be considered as mere
parameters and so \(C\) is a lifting of one of the operators in the
claimed obstructing set. Thus \(C(\Gamma)\) vanishes if all the operators in
the indicated set vanish.
Q.E.D

The properties of the Lie bracket under canonical decomposition are also
quite complex. One can however obviously state
\begin{equation}\label{eq:djbrak}
d_j[F,G] = d_j[d_1F+\cdots+d_jF, d_1G+\cdots+d_jG].
\end{equation}
Concerning symmetries we
begin with some general considerations. Equation~(\ref{eq:symmbrak}) is
equivalent to:
\begin{equation}\label{eq:djsymmbrak}
\hbar\frac{\partial d_jK(t)}{\partial t} = d_j[\ibar F(t), K(t)]
-d_j\frac{\partial}{\partial t}(\tau(t)\ibar F(t))
\end{equation}
for \(j=1,2,\dots\)\@
Let now \(G_\ell(t)\) be a canonical generator of \(K(t)\) at threshold
\(\ell\). From (\ref{eq:djbrak}) and (\ref{eq:djsymmbrak}) one can deduce:
\begin{equation}\label{eq:gensymmbrak}
\hbar\frac{\partial G_\ell(t)}{\partial t} = [\ibar F_\ell(t), L_\ell(t)] +
[\ibar F_\ell(t),G_\ell(t)] + \frac{\partial}{\partial t}(\tau(t)\ibar F_\ell(t))
\end{equation}
where \(L_\ell(t)\) is an operator constructed from the canonical generators of
\(K(t)\) at thresholds less then \(\ell\). Equation
(\ref{eq:gensymmbrak}) is then an inhomogeneous linear evolution equation
for \(G_\ell(t)\) and so in principle can be solved once the generators at
thresholds less then \(\ell\) are known. Thus, barring other
constraints, one can solve
(\ref{eq:symmbrak}) iteratively threshold by threshold. In particular,
if one is given a partial hierarchy \(K(t)\) with
\(n\)-particle operators for \(n \leq \ell\), and if each \(K_n(t)\) is
an infinitesimal symmetry of \(F_n(t)\), then one can, again barring
other constraints, extend the
partial hierarchy to a full symmetry of \(F(t)\) introducing new
generators at thresholds above \(\ell\).

This general statement must be tempered by two considerations if, as is
usually the case, we want symmetries satisfying additional conditions.
In the first place one could at any threshold run into a true
obstruction to extending the symmetry maintaining the additional
conditions, and secondly, given that the extension at any threshold may
not be unique  the existence of such an obstruction can depend on the
choices made at lower thresholds. This means that for practical
calculations, such as those done by computer algebra systems, one should
either always determine the most general solution at each threshold or
be ready to do some backtracking to revise decisions made at lower
thresholds.

We now investigate two specific situations. Generally one would start
with one-particle operators \(F(t)\) and corresponding one-particle
infinitesimal symmetries \(K\). For \(K\) defined by (\ref{eq:InfGen})
we define the canonical lifting \(K^\#\) as  the hierarchy of operators
defined again through (\ref{eq:InfGen}) by \(K(t)^\#\) and the same
function \(\tau(t)\). We can now ask under what conditions is \(K^\#\) a
symmetry of \(F(t)^\#\). Another natural question occurs if now at
threshold \(\ell\) we add new canonical generators \(G(t)\) to \(F(t)\).
We can then ask under what conditions does \(K^\#\) continue being a
symmetry now of \(F(t)^\# + G(t)^\#\). Theorem~\ref{thm:obsatl} provides
simple and useful answers to both questions.

\begin{corollary}\label{cor:onepsymlift}
Let \(F(t)\) and \(K\) be one-particle generators, and
suppose \(K\) is an infinitesimal symmetry of \(F(t)\). The canonical
lifting \(K^\#\), defined in the previous paragraph, is a symmetry of
\(F(t)^\#\) if and only if the two-particle operator \(K^\#_2\) is a
symmetry of \(F(t)_2^\#\) and this happens if and only if  the
two-particle operator

\begin{equation}\label{eq:twopsymmobs}
[F(t)^{\natural(1)},K(t)^{\natural(2)}]+
[F(t)^{\natural(2)},K(t)^{\natural(1)}]
\end{equation}
vanishes.
\end{corollary}

{\em Proof:\/} Since lifting is linear, one has from (\ref{eq:symmbrak})
for the one-particle equations that:
\[
\hbar\frac{\partial K(t)^\#}{\partial t} =
 [\ibar F(t), K(t)]^\# - \frac{\partial}{\partial t}(\tau(t)\ibar
 F(t)^\#).
\]
Thus \(K^\#\) is a symmetry if and only if \([\ibar F(t),
K(t)]^\# = [\ibar F(t)^\#, K(t)^\#]\) and now we apply
Theorem~\ref{thm:obsatl}.

Thus if a symmetry extends to two particles it extends to any number of
particles. One notices that for two real-linear one-particle operators
\(F\) and \(G\), the two-particle operator \([F^{(1)},G^{(2)}]\) is
always zero
as \(F^{(1)}\) and \(G^{(2)}\) act on different tensor factors and
such real-linear operators commute. Thus obstruction
(\ref{eq:twopsymmobs})  can be non-zero only in non real-linear theories.
Nevertheless, for some of the usual symmetries considered, such as
space-time symmetries for particles without internal degrees of freedom,
the obstruction generally vanishes even in the non-linear case. We shall
see this in Section~\ref{sec:spacetime}

\begin{corollary}\label{cor:addgensymm}
Let \(F(t)\) be a tensor derivation and \(K\) a symmetry with only
a one-particle generator: \(K(t) = d_1K(t)\). Let \(G(t)\) be a
\(\ell\)-particle generator with \(\ell > 1\), then \(K\) is a symmetry of \(F(t)+
G(t)^\#\) if and only if \(K_{\ell}\) and \(K_{\ell +1}\) are
symmetries of \(F_{\ell}(t)+G(t)\) and \(F_{\ell+1}(t)+
G(t)_{\ell+1}^\#\) respectively, and this happens if and only if
\([G(t),K_{\ell}(t)] =0 \) and  the following \((\ell
+1)\)-particle operator vanishes:
\begin{equation}\label{eq:addgenobst}
\sum_{j=1}^{\ell+1}[ G(t)^{\jmath^\sharp},
K(t)^{\natural(j)}]
\end{equation}
where  \(\jmath^\sharp \) is the \(\ell\)-tuple.
\((1,\dots,j-1,j+1,\dots,\ell+1)\).
\end{corollary}
{\em Proof:\/} Suppose \(K\) is a symmetry of \(F(t) + G(t)^\#\),
then at particle numbers \(n \geq \ell\) one has from the fact that
\(K\) is a symmetry of \(F(t)\) and (\ref{eq:symmbrak}):
\[
 [\ibar G(t)_n^\#, K_n(t)] - \frac{\partial}{\partial t}(\tau(t)\ibar
 G(t)_n^\#)=0.
\]
Again, since lifting is linear, this equation at
particle number \(\ell\) then implies
\[
 [\ibar G(t), K_{\ell}(t)]_n^\# - \frac{\partial}{\partial t}(\tau(t)\ibar
 G(t)_n^\#)=0
\]
and once again we apply Theorem~\ref{thm:obsatl}.

Thus, under the hypotheses of Corollary~\ref{cor:addgensymm}, if at some
particle number we've added a new generator preserving the symmetry at
that particle number, the symmetry is then preserved in the whole
hierarchy if and only if it is preserved at the next higher particle
number. As before, in the real-linear case obstruction
(\ref{eq:addgenobst}) vanishes automatically.

Theorem~\ref{thm:obsatl} or course covers situation more complex than the ones
covered by the two corollaries above, but the corollaries take care of the most
frequent cases. In the more general situations the obstructions do not
necessarily vanish even in the real-linear case.

\section{Space-time symmetries}\label{sec:spacetime}

In this section we shall consider particles without internal degrees of
freedom. One can admit any number of species, but we suppress any
indication of these, as the reintroduction of species labels is a
straightforward matter as explained in the paragraph following
Definition~\ref{def:hierarchy}. Most of the literature on non-linear
Schr\"odinger equations considers, as a simplifying assumption, only
this type of particle.

By a {\em space-time symmetry\/} \(V(t)\) we mean one related to an underlying transformation
 of space-time
\(\Phi : (t, x) \mapsto (T(t,
x), X(t, x))\).  Our interpretation of such a symmetry is
that it describes the same physical system seen from a
changed reference frame, said change deriving from
\(\Phi\). Our first simplifying assumption is that \(T\)
depends only on \(t\) as otherwise given \(n\) particle
positions \(x_1, \dots ,x_n\) at instant \(t\), the transformed
time instances \(T(t, x_i)\) could all be different and the
construction of a multi-particle wave-function at one time
instant would not be a straightforward matter. We have thus
opted for at most Galileian relativity, if
any, and we have:
\begin{equation}\label{eq:spacetimetrans}
\Phi(t,x) = (T(t), X(t, x)).
\end{equation}

Since the transformed function in principle describes the same physical
system as seen from the transformed frame, the probability densities,
which in principle are objective observable quantities, should transform
accordingly. Likewise it is natural that the symmetry be separating and
we continue to assume this. Finally one must make a conventional choice
deciding if \(\Phi\) describes the old coordinates in terms of the new
or vice-versa. We opt for the old in terms of the new, and thus assume
for the probability densities that:
\begin{equation}\label{eq:probrel}
|V\psi|^2 (t, x_1,\dots,x_n) =
|\psi|^2(T(t),X_1,\dots,X_n)|JX_1|\cdots|JX_n|
\end{equation}
where we've used the abbreviation \(X_i = X(t, x_i)\), and
where \(JX_i\) is the jacobian determinant evaluated at \(x_i\) of the
transformation \(x \mapsto X(t, x)\). For particles with internal
degrees of freedom one may need to sum both sides of (\ref{eq:probrel})
over internal indices. This would modify substantially the rest of the
argument.
From (\ref{eq:VoftT}) we identify the right-hand side as
\(|V(t)\psi(T(t))|^2(x_1,\dots,x_n)\). Equation~(\ref{eq:probrel}) now imposes conditions on the exponential
indices \((a(t),b(t))\) of \(V(t)\). Substituting \(\psi\) by \(k\psi\)
in the equation one deduces that \(\re a(t)= 1\) and \(\im b(t)=0\).
Putting all this together we get:
\begin{theorem}\label{thm:stsymm}
Using the notation of the previous paragraph, a space-time symmetry
\(V\) of type (\ref{eq:VoftT}) associated to a space-time transformation
(\ref{eq:spacetimetrans})
is given
by:
\begin{equation}\label{eq:stsymm}
V(t)\phi = e^{i\Theta(t)(\phi)}\phi(X_1,\dots,X_n)
|JX_1|^{\frac{1}{2}}\cdots|JX_n|^{\frac{1}{2}}
\end{equation}
where \(\Theta(t)\) is a {\em real\/} operator (produces real-valued
functions) with the homogeneity property:
\begin{equation}\label{eq:zeologhom}
\Theta(t)(k\phi) = \Theta(t)(\phi) + (\alpha(t), \beta(t))\cdot \ln k
\end{equation}
for some  real functions \(\alpha(t)\) and \(\beta(t)\).
\end{theorem}
Note that what (\ref{eq:zeologhom}) says is that \(\phi \mapsto
\Theta(t)(\phi) \phi\) is mixed-logarithmic homogeneous with real
indices.

A one parameter group \(V(r)\) of space-time symmetries would be
associated to a one-parameter group \(\Phi(r):(t,x) \mapsto (T(t,r),X(t,x,r))\)
of space-time transformations. Setting
\(T(t,r) =  t + r\tau(t) + o(r)\),
\(X(t,x,r) = x+r\xi(t,x)+o(r)\), and
\(\Theta(t,r)  =  I + r\theta(t) + o(r)\),
one deduces from (\ref{eq:stsymm}):
\begin{theorem}
Using the notation of the previous paragraph,
the infinitesimal generator \(K\) of a space-time symmetry has the form:
\[
K(t)(\phi) = i(\theta(t)\phi)\,\phi + \sum_{j=1}^n\left((\xi\cdot\nabla)^{(j)}
+ \frac{1}{2}(\nabla\cdot\xi)^{(j)}\right)\phi,
\]
where \(\theta(t)\)
is a  real operator such that \(\phi \mapsto \theta(t)(\phi)\phi\) is
mixed-logarithmic homogeneous with real logarithmic indices.
\end{theorem}

The most common form of space-time symmetries are {\em point
symmetries\/} of the type generally considered for differential
equations \cite{Olver}. By
this we mean that \((V(t)\phi)(x) = H(\phi(X(t,x)),t,x)\)
for some complex
function \(H\). This amounts to saying that the transformation
\((w,t,x) \mapsto (H(w,\Phi^{-1}(t,x)),\Phi^{-1}(t,x))\) maps the graph of
\(\psi\) to the graph of \(V\psi\); that is, the symmetry is effected by
a point transformation in the space in which the graph of a solution
lies. One can envisage more general transformations in which \(H\) is a
differential operator \cite{OS} and though there are differential
equations with such generalized symmetries we shall not pursue this
here.

The homogeneity and separation property of \(V(t)\) now imposes a condition on the
function \(H\). In fact, from \(V(t)(k\phi) =
k^{(1+i\alpha(t),\beta(t))}V(t)(\phi)\) one deduces \(H(kw,t,x)=
k^{(1+i\alpha(t),\beta(t))}H(w,t,x)\) from which setting \(w=1\) and renaming
\(k\) as \(w\) one gets
\(H(w,t,x) = w^{(1+i\alpha(t),\beta(t))}H(1,t,x)\) and the separation
property implies \(H(1,t,x)= \prod_{j=1}^nN(t,x_j)\)
for some complex function \(N\). Comparing this with (\ref{eq:stsymm})
and then also deducing the version for infinitesimal symmetries, one
finds:
\begin{theorem}
A point space-time symmetry, using the notation of the paragraph
preceding Theorem~\ref{thm:stsymm}, has the form:
\[
V(t)\phi = e^{i\left(\sum_{j=1}^n\upsilon^{(j)}(t)\right)}(\phi(X_1,\dots,X_n))^{(1+i\alpha(t),\beta(t))}|JX_1|^{\frac{1}{2}}\cdots|JX_n|^{\frac{1}{2}}
\]
for some real functions \(\alpha(t)\) and \(\beta(t)\) and \(\upsilon(t,x)\). An infinitesimal
point space-time symmetry has the form:
\begin{equation}\label{eq:pstisymm}
K(t)\phi = \sum_{j=1}^n\left(i\eta^{(j)}+ (\xi\cdot\nabla)^{(j)} +
\frac{1}{2}(\nabla\cdot\xi)^{(j)}\right)\phi + i(\gamma(t),\delta(t))\cdot\ln \phi\,\phi
\end{equation}
for some real functions \(\gamma(t)\) and \(\delta(t)\) and \(\eta(t,x)\).
\end{theorem}
We see from (\ref{eq:pstisymm}) that an infinitesimal point space-time
symmetry is always a canonical lift from the one-particle generator.

\begin{theorem}\label{thm:freelift}
The lifting obstructions (\ref{eq:twopsymmobs}) and
(\ref{eq:addgenobst}) vanish for point space-time symmetries.
\end{theorem}
{\em Proof:\/} We see from (\ref{eq:pstisymm}) that for the one-particle
symmetry\[K(t)^\natural =
i\eta\phi + \xi\cdot\nabla\phi +
\frac{1}{2}(\nabla\cdot\xi)\phi\] which is a first order linear
differential operator. Since the obstruction
expressions are
real-linear in \(K(t)^\natural\) we can consider each term separately.
To expedite the argument one can in an obvious manner consider
(\ref{eq:twopsymmobs}) as the \(\ell=1\) case of (\ref{eq:addgenobst}),
remembering that \(G^\natural = G\) for a generator at particle number
above one. For a strictly homogeneous one-particle operator \(L\)
consider one term in the obstruction (\ref{eq:addgenobst}):
\begin{equation}\label{eq:ob}
[ G^{\jmath^\sharp}, L^{(j)}].
\end{equation}
Let \(\alpha(t,x)\) be a real one-particle function and consider
\(\alpha L\). One has \[\alpha^{(j)}\Fd
G^{\jmath^\sharp}(\phi)\cdot L^{(j)}(\phi)  = \Fd
G^{\jmath^\sharp}(\phi)\cdot \alpha^{(j)}L^{(j)}(\phi)\]
since \(\Fd G^{\jmath^\sharp}\) is real-linear and the position variables upon
which it acts are disjoint from the position variable in
\(\alpha^{(j)}\) allowing us to consider this function  as a parameterized real
number. One therefore has \([ G^{\jmath^\sharp},\alpha^{(j)} L^{(j)}] =
\alpha^{(j)}[ G^{\jmath^\sharp}, L^{(j)}]\). From this, to prove the theorem one need only show that
(\ref{eq:ob}) vanishes for \(L\) equal to the identity \(I\), to
\(iI\),
and to a partial derivative. Now for any complex number \(\eta\),
\(\Fd\eta I = \eta I\), and so
Equation (\ref{eq:Eulerpow}) for a strictly homogeneous operator states
exactly that the Lie bracket of that operator with \(\eta I\) vanishes.
This takes care of the first two cases.
Let now \(L\) be a partial derivative
\({\partial}/{\partial x^k}\). Since \(L^{(j)}\) acts on what
to \(G^{\jmath^\sharp}\) are just parameters, we have by the chain rule for
Fr\'echet derivatives:\[L^{(j)}G^{\jmath^\sharp}\phi = \Fd
G^{\jmath^\sharp}\cdot L^{(j)}\phi.\] Again, since
\(\Fd L = L\), the above equation just says
that (\ref{eq:ob}) vanishes. Q.E.D.

We see therefore that although there are true obstructions to lifting of
symmetries in non-linear theories, point space-time symmetries in
separating hierarchies for particles without internal degrees of freedom
are not affected by these. This result justifies the claims made about
the Galileian invariance of the modified Doebner-Goldin hierarchy at the
end of \cite{GS:separation}.

Some of the cases covered by the last theorem are otherwise obvious for
hierarchies of differential operators. For instance such operators are
symmetric under space translations if each operator does not depend
explicitly on
the spatial coordinates, and this property obviously is maintained by
liftings. For other symmetries such as Galileian boosts, or for
operators that are not differential, one must rely on the theorem.

The case of particles with internal degrees
of freedom is different and the possible non-vanishing of the obstructions
can be viewed as either new phenomena, or the necessity of further
constraints upon possible theories requiring that the obstructions
vanish, or the necessity of introducing new generators to maintain
symmetry. This shows that the simplifying assumption of particles without
internal degrees of freedom is not as innocuous a postulate as it may
seem to be.

To understand the problem involved suppose we wish to construct a
hierarchy of equations (\ref{eq:hierarchy}) where the particles have
internal degrees of freedom, which we indicate in the usual way by
putting indices on the wave-function (rather than adding components to
the position variable) assuming that such objects transform according to
some representation (which may be non-linear) of an appropriate symmetry
group. Assume the one-particle equation are symmetric under this group
and write the action of the one-particle operator as
\[F_\alpha(\phi_{\cdot})\] where \(\alpha\) is the internal degree of
freedom index and where we've indicated by the subscript ``\(\cdot\)" on
the wave-function the fact that the operator \(F\) acts on these
components of \(\phi\) to produce in the end another object with the same
transformational properties under the symmetry group. The canonically
lifted two-particle operator can now be written as:
\[F^{(1)}_\alpha(\phi_{\cdot\beta}) +
F^{(2)}_\beta(\phi_{\alpha\cdot}).\] Now in the first term \(\beta\) is
just a parameter and in the second term \(\alpha\) is. For non-linear
\(F\) the above object does not in general transform appropriately in the
pair of indices \((\alpha,\beta)\) and so the canonically lifted
two-particle equation is not symmetric. The only way to recover symmetry
is to introduce two-particle generators either in the equations (which
is the more likely) or the infinitesimal symmetry.  But the same
problem will then arise at three-particles and by repeating the argument
one sees that in general one will need to introduce new generators at
each particle number. Thus in contrast to the case of no internal
degrees of freedom where one can construct symmetric hierarchies with
just a finite number of generators, for theories with internal degrees
of freedom one needs an infinite number. For such theories to be
tenable, one must either abandon the symmetry at some particle number
or introduce some principle that would systematically pick out the
needed generators.

The problem is most acute for theories with spin.  One imposes rotation
invariance on the grounds that space itself is isotropic. In linear
theories, once the one-particle equation is chosen and has the
appropriate transformation property with respect to the rotation group,
then the multi-particle equations are unique and automatically have the
right (tensorial) transformation properties. In the non-linear case one
must assert or deny rotation invariance for each particle number. To
deny it at any point would call in question the very idea of space
isotropy, and to assert it universally one must then make infinite
choices of generators for the hierarchy of equations. This is another
challenge that non-linear theories must meet.

\subsection{Acknowledgment}
The author thanks professor Gerald Goldin for his interest, motivation
and helpful discussions. Special thanks go to the Mathematics Department
of Rutgers University for its hospitality during the author's stay
there. This research was financially supported by the Secretaria de
Ci\^encia e Tecnologia (SCT) and the Conselho Nacional de
Desenvolvimento Cient\'{\i}fico e Tecnol\'ogico (CNPq), both agencies of
the Brazilian government.

\end{document}